\documentclass[10pt,jkps,preprint,fleqn,showpacs,showkeys,twocolumn]{revtex4}
\usepackage{graphicx}
\usepackage{amssymb}
\usepackage{amsmath}
\usepackage{bm}
\begin{document}
\setcounter{page}{1}
\title[]{Arbitrary amplitude ion acoustic solitary structures in a collisionless magnetized plasma consisting of nonthermal and isothermal electrons}
\author{Sandip \surname{Dalui}}
\email{dalui.sandip77@gmail.com}
\affiliation{Department of Mathematics, Techno Main Salt Lake, Kolkata - 700091, India.}
\affiliation{Department of Mathematics, Jadavpur University, Kolkata - 700032, India.}
\author{Sankirtan \surname{Sardar}}
\affiliation{Department of Mathematics, Jadavpur University, Kolkata - 700032, India.}
\affiliation{Advanced center for non-linear and complex phenomena, Kolkata  - 700075, India.}
\author{Anup \surname{Bandyopadhyay}}
\email{abandyopadhyay1965@gmail.com}
\affiliation{Department of Mathematics, Jadavpur University, Kolkata - 700032, India.}
\affiliation{Advanced center for non-linear and complex phenomena, Kolkata  - 700075, India.}
%\thanks{Fax: +82-2-554-1643}

\date[]{Received \today}

\begin{abstract}
We have used the Sagdeev pseudo-potential technique to investigate the arbitrary amplitude ion acoustic solitons, double layers and supersolitons in a collisionless magnetized plasma consisting of adiabatic warm ions, isothermal cold electrons and nonthermal hot electrons immersed in an external uniform static magnetic field. We have used the phase portraits of the dynamical system describing the nonlinear behaviour of ion acoustic waves to confirm the existence of different solitary structures. We have also investigated the transition of different solitary structures: soliton (before the formation of double layer) $\rightarrow$ double layer $\rightarrow$ supersoliton $\rightarrow$ soliton (soliton after the formation of double layer) by considering the variation of $\theta$ only, where $\theta$ is the angle between the direction of the external uniform static magnetic field and the direction of propagation of the wave.
\end{abstract}


\maketitle

\section{INTRODUCTION}
Sagdeev \cite{sagdeev1966reviews} introduced a comprehensive nonlinear method for arbitrary amplitude ion acoustic (IA) solitary waves in the literature. Washimi \& Taniuti \cite{washimi1966propagation} investigated the small amplitude IA solitary waves in a collisionless plasma composed of cold ions and hot isothermal electrons. Yu \textit{et al.} \cite{yu1980fully} investigated arbitrary amplitude IA solitary waves in a magnetized plasma consisting of cold ions and isothermal electrons. Choi \textit{et al.} \cite{choi2005ion} investigated arbitrary amplitude IA solitary waves in a dusty plasma obliquely propagating to an external magnetic field. In a very recent paper, Dalui \textit{et al.} \cite{dalui2017modulational_2} have investigated modulational instability of small amplitude IA waves in a collisionless magnetized plasma consisting of warm adiabatic ions, Maxwell-Boltzmann distributed cold electrons and Cairns \cite{cairns1995electrostatic} distributed nonthermal hot electrons immersed in a uniform static magnetic field ($\textbf{B} = B_{0}\hat{\textbf{z}}$) propagating along $z-$axis. Rufai \textit{et al.} \cite{rufai2014ion} investigated the arbitrary amplitude IA solitary waves in a magnetized plasma consisting of cold ions, Maxwell-Boltzmann distributed cold electrons and Cairns \cite{cairns1995electrostatic} distributed nonthermal hot electrons. Rufai \textit{et al.} \cite{rufai2016nonlinear} investigated the finite amplitude IA solitary waves in a collisionless magnetized plasma consisting of adiabatic warm ions and Cairns \cite{cairns1995electrostatic} distributed nonthermal hot electrons. In the present paper, we have extended this paper of Rufai \textit{et al.} \cite{rufai2016nonlinear} in the following directions: 
\begin{itemize}
	\item (i) instead of considering only one electron species, two different species of electrons at different temperatures have been considered,
	\item (ii) a one - to - one correspondence between the phase portraits describing the nonlinear behaviour of ion acoustic waves and the graph of $V(\phi)$ against $\phi$ has been established, where $V(\phi)$ is the Sagdeev pseudo-potential with $\phi$ is the electrostatic potential,
	\item (iii) the smooth transition of different solitary structures, viz., soliton (before the formation of double layer) $\rightarrow$ double layer $\rightarrow$ supersoliton $\rightarrow$ soliton (soliton after the formation of double layer), has been investigated. 
\end{itemize}

Therefore, the present paper can be regarded as a new problem with respect to the above-mentioned considerations.
\section{BASIC EQUATIONS}
In the present paper, we have studied the arbitrary amplitude ion acoustic solitary waves, double layers and supersoliton structures by considering exactly the same plasma system of Dalui \textit{et al.} \cite{dalui2017modulational_2} but here we have assumed the quasi neutrality condition of charged particulates instead of considering Poisson equation on the basis of the assumption that the length scale of the solitary wave is greater than the Debye length as well as the ion gyroradius \cite{choi2005ion}. So, here we have considered a collisionless plasma consisting of warm adiabatic ions, Maxwell-Boltzmann distributed cold isothermal electrons and Cairns \cite{cairns1995electrostatic} distributed nonthermal hot electrons immersed in an external uniform static magnetic field ($\textbf{B} = B_{0}\hat{\textbf{z}}$) directed along $z$-axis. The nonlinear behaviour of IA waves can be described by the continuity equation, the equation of motion and the pressure equation of ion fluids together with the quasi neutrality condition:
%---------------------------------------------------------------------------------------------------------------------------------------
\begin{eqnarray}
 && \frac{\partial n}{\partial t} + \nabla \cdot (n\textbf{u}) = 0 , \label{P5_continuity_equation}
\end{eqnarray}
\begin{eqnarray}
&& \frac{\partial \textbf{u}}{\partial t} + (\textbf{u} \cdot \nabla)\textbf{u} + \frac{\sigma}{n}\nabla p + \nabla \phi  =  \omega_{c}(\textbf{u} \times \hat{\textbf{z}}) , \label{P5_Equation_of_motion}
\end{eqnarray}
\begin{eqnarray}
 && p = n^{\gamma} , \label{P5_Pressure_equation}
\end{eqnarray}
\begin{eqnarray}
&& n = n_{ce}+n_{se} , \label{P5_quasi_neutrality} 
\end{eqnarray}
where $\nabla \equiv (\frac{\partial}{\partial x},\frac{\partial}{\partial y},\frac{\partial}{\partial z})$.
%----------------------------------------------------------------------------------------------------------------------------------------------------------
Here we have used the notations $t$ for time, $(x,y,z)$ for spatial variables, $\textbf{u}=(u,v,w)$ for ion fluid velocity vector, $\omega_{c}$ for ion cyclotron frequency, $\phi$ for electrostatic potential. Again, $n_{se}$, $n_{ce}$ and $n$ are number densities of isothermal electron species, nonthermal electron species and ion species respectively.  Here $t$, $(x,y,z)$, $\textbf{u}=(u,v,w)$, $\omega_{c}$, $\phi$, $n_{se}$, $n_{ce}$ and $n$ are normalized variables and these quantities have been normalized by $\omega_{pi}^{-1}[= \sqrt{m/(4\pi n_{0} e^{2})}]$, $\lambda_{Df}$ $(= \sqrt{K_{B} T_{ef}/4\pi n_{0} e^{2}} )$, $c_{s}$ $(= \sqrt{K_{B} T_{ef}/m})$, $\omega_{pi}$, $K_{B} T_{ef}/e$, $n_{0}$, $n_{0}$ and $n_{0}$ respectively, where $\sigma = T_{i}/T_{ef}$, $\gamma =\frac{5}{3}$, $K_{B}$ $=$ Boltzmann constant, $m$ $=$ mass of an ion, $-e$ $=$ charge of an electron, $T_{i}$ $=$ average ion temperature and $T_{ef}$ is given by
\begin{eqnarray}\label{P5_equation_of_effective_temperature}
 \frac{ n_{c0} + n_{s0}}{T_{ef}} =  \frac{ n_{c0}}{T_{ce}} + \frac{ n_{s0}}{T_{se}} ,
\end{eqnarray}
where ${n}_{c0}$, ${n}_{s0}$, $T_{ce}$ and $T_{se}$ are unperturbed nonthermal electron number density, unperturbed isothermal electron number density, average temperature of nonthermal electrons and average temperature of isothermal electrons respectively.

Using the normalization as discussed before for the independent and dependent variables, the number densities of nonthermal and isothermal electrons can be written in the following form:
%-----------------------------------------------------------------------------------------------------------------------------------------------------------------
\begin{eqnarray}
  n_{ce} = \bar{n}_{c0}(1-\beta_{e} \sigma_{c} \phi + \beta_{e} \sigma_{c}^{2} \phi^{2})\exp[\sigma_{c} \phi] ,\label{P5_Electron_distribution_ce}
\end{eqnarray}
\begin{eqnarray}
  n_{se} = \bar{n}_{s0}\exp[\sigma_{s} \phi] ,\label{P5_Electron_distribution_se}
\end{eqnarray}
where $ \beta_{e} = \frac{4 \alpha_{e}}{1+3 \alpha_{e}}$  with $\alpha_{e} \geq 0$, $\bar{n}_{c0}=\frac{n_{c0}}{n_{0}}$, $\bar{n}_{s0}=\frac{n_{s0}}{n_{0}}$, $\sigma_{c} = \frac{T_{ef}}{T_{ce}}$, $\sigma_{s} = \frac{T_{ef}}{T_{se}}$. Here, $\beta_{e}$ is the nonthermal parameter. Now, the equation (\ref{P5_equation_of_effective_temperature}) and the unperturbed charge neutrality condition ($n_{c0} + n_{s0} = n_{0}$) can be written as
\begin{eqnarray}\label{P5_charge_neutrality_condition_1}
  \bar{n}_{c0} \sigma_{c} + \bar{n}_{s0} \sigma_{s} = 1,\bar{n}_{c0} + \bar{n}_{s0} = 1 .
\end{eqnarray}

So, the basic parameters of the present plasma system are $\gamma$, $\sigma$, $\sigma_{sc} = \frac{T_{se}}{T_{ce}}$, $n_{sc} = \frac{n_{s0}}{n_{c0}}$, $\beta_{e}$ and $\omega_{c}$. With respect to the parameters $\sigma_{sc}$ and $n_{sc}$, one can use the following equations:
\begin{eqnarray}\label{P5_form_of_sigma_c1}
\Big(\bar{n}_{s0}~,~\bar{n}_{c0}\Big) =  \frac{1}{1+n_{sc}}\bigg(n_{sc}~,~1\bigg),
\end{eqnarray}
\begin{eqnarray}\label{P5_form_of_sigma_c2}
\Big(\sigma_{s}~,~\sigma_{c} \Big) = \frac{1+n_{sc}}{\sigma_{sc}+n_{sc}}\Big(1~,~\sigma_{sc}\Big),
\end{eqnarray}
where we have used the first and second equations of (\ref{P5_charge_neutrality_condition_1}) to find the expressions $\bar{n}_{s0}$, $\bar{n}_{c0}$, $\sigma_{s}$ and $\sigma_{c}$.

Now using the equation (\ref{P5_Pressure_equation}), the equation (\ref{P5_Equation_of_motion}) can be written as
%-----------------------------------------------------------------------------------------------------------------------------------------------------------
\begin{equation}\label{P5_Equation_of_motion_1}
\frac{\partial \textbf{u}}{\partial t} + (\textbf{u} \cdot \nabla)\textbf{u} + \frac{\sigma \gamma}{\gamma-1} \nabla (n^{\gamma-1}) + \nabla \phi  =  \omega_{c}(\textbf{u} \times \hat{\textbf{z}}) .
\end{equation}
Using (\ref{P5_Electron_distribution_ce}) and (\ref{P5_Electron_distribution_se}), the equation (\ref{P5_quasi_neutrality}) can be written as
\begin{equation}\label{P5_quasi_neutrality_1} 
  n =  \bar{n}_{c0}(1-\beta_{e} \sigma_{c} \phi + \beta_{e} \sigma_{c}^{2} \phi^{2})\exp[\sigma_{c} \phi] 
      + \bar{n}_{s0}\exp[\sigma_{s} \phi] .
\end{equation}

For low frequency IA waves ($\omega<<\omega_{c}$), the linear dispersion relation of IA waves can be easily obtained from the equations (\ref{P5_continuity_equation}), (\ref{P5_Equation_of_motion_1}) and  (\ref{P5_quasi_neutrality_1}) and this dispersion relation is
\begin{eqnarray}\label{P5_dispersion_relation_linear_300}
  \frac{\omega}{k_{\parallel}} = \bigg[\frac{k_{\perp}^{2}}{\omega_{c}^{2}} +  M_{s}^{-2} \bigg]^{-\frac{1}{2}} ,
\end{eqnarray}
where the perturbed dependent variables are assumed to vary as $\exp [i(k_{x}x+k_{y}y+k_{z}z-\omega t)]$. Here, $k_{\perp}^{2} = k_{x}^2 + k_{y}^2$, $k_{\parallel}^{2} = k_{z}^2$ and the expression of $M_{s}$ is given by the following equation
\begin{eqnarray}\label{P5_Ms}
  M_{s} = \sqrt{ \gamma\sigma + \frac{n_{sc}+\sigma_{sc}}{n_{sc}+(1-\beta_{e})\sigma_{sc} } } .
\end{eqnarray}
 
%%%%%%%%%%%%%%%%%%%%%%%%%%%%%%%%%%%%%%%%%%%%%%%%%%%%%%%%%%%%%%%%%%%%%%%%%%%%%%%%%%%%%%%%%%%%%%%%%%%%%%%%%%%%%%%%%%%%%%%%%%%%%%%%%%%%%%%%%%%%%%%%%%%%%%%%%%%%%%%%%%%%%%%%%%%%%%%%%%%%%%%%%%%%%%%%%%%%%%%%%%%%%%%%%%%%%%%%%%%%%%%%%%%%%%%%%%%%%%%%%%%%%%%%%%%%%%%%%%%%%%%%%%%%%%%%%%%%%%%%%%%%%%%%%%%%%%%%%%%%%%%%%%%%%%%%%%%%%%%%%%%%%%%%%%%%%%%%%%%%%%%%%%%%%%%%%%%%%%%%%%%%%%%%%%%%%%%%%%%%%%%%%%%%%%%%%%%%%%%%%%%%%%%%%%%%%%%%%%%%%%%%%%%%%%%%%%%%%%%%%%%%%%%%%%%%%%%%%%%%%%%%%%%%%%%%%%%%%%%%%%%%%%%%%%%%%%%%%%%%%%%%%%%%%%%%%%%%%%%%%%%%%%%%%%%%%%%%%%%%%%%%%%%%%%%%%%%%%%%%%%%%%%%%%%%%%%%%%%%%%%%%%%%%%%%%%%%%%%%%%%%%%%%%%%%%%%%%%%%%%%%%%%%%%%%%%%%%%%%%%%%%%%%%%%%%%%%%%%%%%%%%%%%%%%%%%%%%%%%%%%%%%%%%%%%%%%%%%%%%%%%%%%%%%%%%%%%%%%%%%%%%%%%%%%%%%%%%%%%%%%%%%%%%%%%%%%%%%%%%%%%%%%%%%%%%%%%%%%%%%%%%%%%%%%%%%%%%%%%%%%%%%%%%%%%%%%%%%%%%%%%%%%%%%%%%%%%%%%%%%%%%%%%%%%%%%%%%%%%%%%%%%%%%%%%%%%%%%%%%%%%%%%%%%%%%%%%%%%%%%%%%%%%%%%%%%%%%%%%%%%%%%%%%%%%%%%%%%%%%%%%%%%%%%%%%%%%%%%%%%%%%%%%%%%%%%%%%%%%%%%%%%%%%%%%%%%%%%%%%%%%%%%%%%%%%%%
\section{\label{P5_sec_Energy_integral} Energy Integral}
To study the arbitrary amplitude time independent IA solitary structures, we consider the transformation:
\begin{eqnarray}\label{P5_Transformation}
  \xi=l_{x} x + m_{y} y + n_{z} z - M t ,
\end{eqnarray}
where $l_{x}^{2} + m_{y}^{2} + n_{z}^{2} = 1$ and $M$ is normalized velocity of the wave frame known as Mach number. Here $M$ is normalized by $c_{s}$. 

Therefore, transforming the equation (\ref{P5_continuity_equation}) and each component of the momentum equation (\ref{P5_Equation_of_motion_1}) in the wave frame, which is moving with a constant velocity $M$ along a direction $l_{x} \hat{x}+m_{y}\hat{y}+n_{z}\hat{z}$, we get the following equations:
\begin{eqnarray}
&& \frac{d }{d \xi}(n\Psi) = 0  , \label{P5_Continuity_equation_2}
\end{eqnarray}
\begin{eqnarray}
&& \Psi \frac{d u}{d \xi}+ l_{x}\frac{d H}{d \xi}=\omega_{c}v   ,  \label{P5_x_component_of_equation_of_motion_2}
\end{eqnarray}
\begin{eqnarray}
&& \Psi \frac{d v}{d \xi}+ m_{y}\frac{d H}{d \xi}=-\omega_{c}u  , \label{P5_y_component_of_equation_of_motion_2}
\end{eqnarray}
\begin{eqnarray}
&& \Psi \frac{d w}{d \xi}+ n_{z}\frac{d H}{d \xi}=0   , \label{P5_z_component_of_equation_of_motion_2}
\end{eqnarray}
where 
\begin{eqnarray}
 && \Psi=l_{x} u + m_{y} v + n_{z} w - M   , \label{P5_Form_of_psi} 
 \end{eqnarray}
\begin{eqnarray}
 && H = \phi + \frac{\sigma \gamma }{\gamma-1} n^{(\gamma-1)} . \label{P5_Form_of_H}
\end{eqnarray}
%----------------------------------------------------------------------------------------------------------------------------------------------------------
Integrating the equation (\ref{P5_Continuity_equation_2}) with respect to $\xi$, we get
\begin{equation}\label{P5_Form_of_psi_1}
 \Psi= - \frac{M}{n} ,
\end{equation}
where we have used the boundary conditions:
\begin{equation}\label{P5_bcs}
 (n,u,v,w,\phi,\frac{d \phi}{d \xi}) \rightarrow (1,0,0,0,0,0) \mbox{    as    $|\xi| \rightarrow \infty$}.
\end{equation}
Again, integrating the equation (\ref{P5_z_component_of_equation_of_motion_2}) with respect to $\xi$, we get the expression of $w$ as
\begin{equation}\label{P5_w}
 w = \frac{n_{z}}{M} G , 
\end{equation}
where we have used the same boundary conditions (\ref{P5_bcs}) and $G$ is given by the following equation:
\begin{eqnarray}\label{P5_G_phi}
  G &=& \sigma (n^{\gamma}-1) + \frac{\bar{n}_{s0}}{\sigma_{s}} \exp [\sigma_{s}\phi] - \frac{\bar{n}_{s0}}{\sigma_{s}}  \nonumber \\
    &+& \frac{\bar{n}_{c0}}{\sigma_{c}} \big(1 + 3\beta_{e} - 3\beta_{e}\sigma_{c}\phi + \beta_{e}\sigma_{c}^{2}\phi^{2} \big) \exp [\sigma_{c}\phi] \nonumber \\ 
    &-& \frac{\bar{n}_{c0}}{\sigma_{c}} (1+3\beta_{e}) . 
\end{eqnarray}

Solving the equations (\ref{P5_x_component_of_equation_of_motion_2}) and (\ref{P5_y_component_of_equation_of_motion_2}), the expressions of $u$ and $v$ can be written in the following form:
\begin{eqnarray}
 u = \frac{1}{l_{x}^{2}+m_{y}^{2}} \Bigg[ l_{x}  \Big ( M - \frac{M}{n} - \frac{n_{z}^{2}}{M} G  \Big ) - \frac{m_{y}}{\omega_{c}} \frac{d S}{d \xi} \Bigg] , \label{P5_u} 
 \end{eqnarray}
\begin{eqnarray}
 v = \frac{1}{l_{x}^{2}+m_{y}^{2}} \Bigg[ m_{y}  \Big ( M - \frac{M}{n} - \frac{n_{z}^{2}}{M} G  \Big ) + \frac{l_{x}}{\omega_{c}} \frac{d S}{d \xi} \Bigg] , \label{P5_v}
\end{eqnarray}
where 
\begin{eqnarray}\label{P5_S}
 S = \phi + \frac{\sigma \gamma }{\gamma-1} n^{\gamma-1} + \frac{M^2}{2n^{2}} . 
\end{eqnarray}

Substituting the expressions of $u$ and $v$ in equations (\ref{P5_x_component_of_equation_of_motion_2}) and (\ref{P5_y_component_of_equation_of_motion_2}), we get the following equation:
\begin{eqnarray}\label{P5_S_ode}
 \frac{d^{2} S}{d\xi^{2}} = \omega_{c}^{2} F(\phi)  , 
\end{eqnarray}
where
\begin{eqnarray}\label{P5_S_F_phi}
 F(\phi) = \Big[ n - 1 - \frac{n_{z}^{2}}{M^2} n G \Big]  . 
\end{eqnarray}
Finally, using simple algebra along with the boundary conditions as given in (\ref{P5_bcs}), we get the following energy integral from equation (\ref{P5_S_ode}):

\begin{equation}\label{P5_energy_integral}
 \frac{1}{2} \bigg(\frac{d \phi}{d \xi} \bigg)^{2} + V(\phi) =0  , 
\end{equation}
where
\begin{equation}\label{P5_V_phi}
  V(\phi) = - \omega_{c}^{2} \frac{\int_{0}^{\phi}\frac{d S}{d \phi}F(\phi)d\phi}{\Big(\frac{d S}{d \phi} \Big)^{2}}  .
\end{equation}

Equation (\ref{P5_energy_integral}) is the well-known energy integral with Sagdeev pseudo-potential $V(\phi) = V(\phi,M)$ given  in  (\ref{P5_V_phi}).

As discussed by several authors \cite{sagdeev1966reviews,rufai2014ion,debnath2018ion} the positive (negative) potential solitary wave solutions exist if the conditions are satisfied: (i) $V(\phi,M)=0=V'(\phi,M)$ and $V''(\phi,M)<0$ at $\phi=0$; (ii) $V(\phi_{m},M)=0$ and $V'(\phi_{m},M)>0 (V'(\phi_{m},M)<0)$ for some $\phi_{m}>0(\phi_{m}<0)$; (iii) $V(\phi)<0$ for all min$\{0,\phi_{m} \}$ $<\phi<$ max$\{0,\phi_{m} \}$. Also for the existence of positive (negative) potential double layer solution of (\ref{P5_energy_integral}) we must replace the second condition by $V(\phi_{m},M)=0$, $V'(\phi_{m},M)=0$ and $V''(\phi_{m},M)<0$ for some $\phi_{m}>0(\phi_{m}<0)$.  

\section{SOLITARY STRUCTURES and PHASE PORTRAITS}
Differentiating the energy integral (\ref{P5_energy_integral}) with respect to $\phi$, we get 
\begin{eqnarray}\label{P5_d_phi_d_xi}
 \frac{d^{2} \phi}{d \xi^{2}}  + V'(\phi) =0  .
\end{eqnarray}
The equation (\ref{P5_d_phi_d_xi}) is equivalent to the following system of differential equations:
\begin{eqnarray}\label{P5_phase_PORTRAITS}
 \frac{d \phi_{1}}{d \xi}  =\phi_{2} ~, ~~     \frac{d \phi_{2}}{d \xi}   = - V'(\phi_{1})  .
\end{eqnarray}

%----- FIG-1 --------------
\begin{figure}[ht]
\begin{center}
\includegraphics{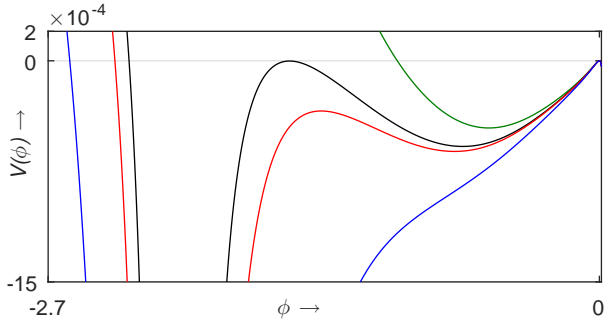}
  \caption{\label{P5_soliton_double_layer_super_soliton_100000} $V(\phi)$ is plotted against $\phi$ for $\sigma=0.001$, $\gamma=5/3$, $M=0.98$, $n_{sc}=1/9$, $\sigma_{sc}=0.04$, $\beta_{e}=4/103$ and $\omega_{c}=0.2$. Green curve corresponds to $\cos 39^{\circ}$, black curve corresponds to $\cos 39.752^{\circ}$, red curve corresponds to $\cos 39.9^{\circ}$ and blue curve corresponds to $\cos 41^{\circ}$.}
\end{center}
\end{figure}
In figure \ref{P5_soliton_double_layer_super_soliton_100000}, $V(\phi)$ has been plotted against $\phi$ for different values of $n_{z}$ ($=\cos \theta$). The green curve represents soliton structure, the black curve represents double layer structure, the red curve represents supersoliton structure and the blue curve represents soliton structure (soliton after the formation of double layer). In this figure, we have shown the formation of different solitary structures, viz., soliton (soliton before the formation of double layer), double layer, supersoliton, soliton (soliton after the formation of double layer) for different values of $n_{z}$ and fixed values of the other parameters of the system and also for the fixed value of $M$.  

%----- FIG-2 --------------
\begin{figure}[ht]
\begin{center}
\includegraphics{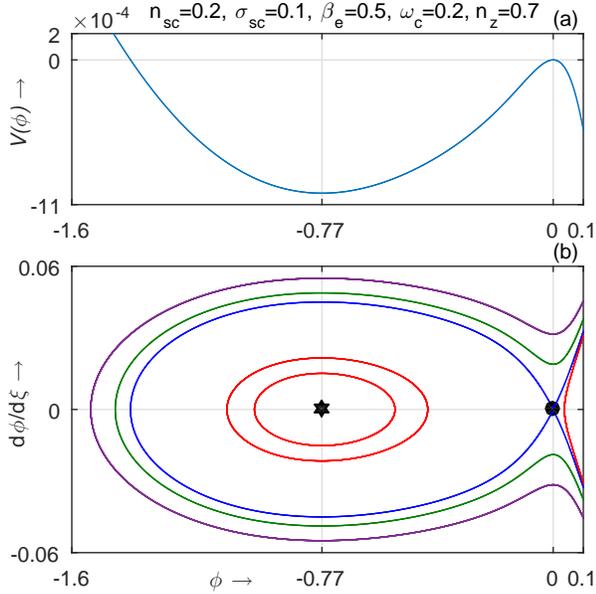}
  \caption{\label{P_5_negative_soliton_phi_vs_vphi} $V(\phi)$ and $\frac{d\phi}{d\xi}$ are plotted against $\phi$ in (a) and (b), respectively, for $\sigma=0.001$, $\gamma=5/3$ and $M=0.98$. We see that the curve $V(\phi)$ of figure \ref{P_5_negative_soliton_phi_vs_vphi}(a) corresponds to the separatrix of the phase portrait as shown by the blue curve of figure \ref{P_5_negative_soliton_phi_vs_vphi}(b). The maximum value of $V(\phi)$ generates a saddle point whereas the minimum value of $V(\phi)$ generates an another equilibrium point (non-saddle fixed point) of the dynamical system (\ref{P5_phase_PORTRAITS}). This figure represents the formation of the negative potential solitary wave.}
\end{center}
\end{figure}

%----- FIG-3 --------------
\begin{figure}[ht]
\begin{center}
\includegraphics{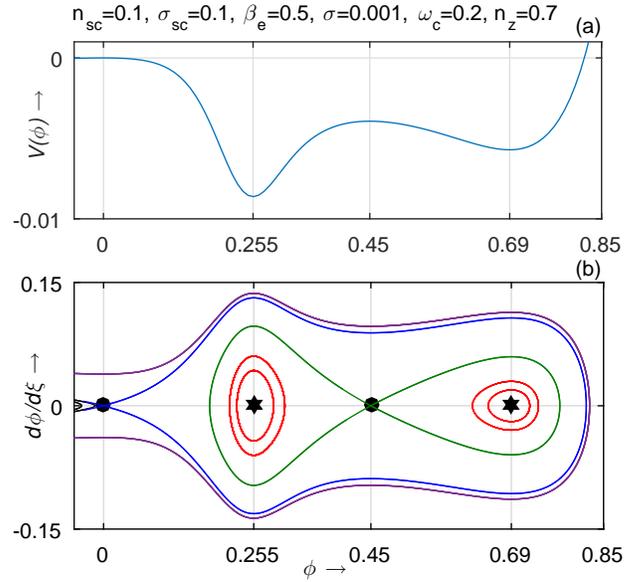}
  \caption{\label{P5_Supersoliton_2} $V(\phi)$ and $\frac{d\phi}{d\xi}$ are plotted against $\phi$ in (a) and (b), respectively, for $\sigma=0.001$, $\gamma=5/3$ and $M=0.98$. Here, we see that the separatrix of the phase portrait as shown by the blue curve of figure \ref{P5_Supersoliton_2}(b) which appears to pass through the saddle $(0,~0)$ point contains two non-saddle fixed points and another separatrix (green curve) appears to pass through the non-zero saddle $(0.45,~ 0)$ point. Finally, the separatrix (blue curve) that appears to pass through the saddle $(0,~ 0)$ point represents a supersoliton. Therefore, this figure represents the formation of supersoliton.}
\end{center}
\end{figure}

In figure \ref{P_5_negative_soliton_phi_vs_vphi},  $V(\phi)$ and $\frac{d\phi}{d\xi}$ are plotted against $\phi$ in \ref{P_5_negative_soliton_phi_vs_vphi}(a) and \ref{P_5_negative_soliton_phi_vs_vphi}(b) respectively. Figure \ref{P_5_negative_soliton_phi_vs_vphi}(a) confirms the existence of negative potential solitary wave. Corresponding to this negative potential solitary wave in figure \ref{P_5_negative_soliton_phi_vs_vphi}(b), we have shown a phase portrait of the system (\ref{P5_phase_PORTRAITS}). The small solid circle at the point $(0, 0)$ corresponds to a saddle point whereas the small solid hexagon at the point $(-0.77, 0)$ indicates a stable equilibrium point of the system (\ref{P5_phase_PORTRAITS}). From figure \ref{P_5_negative_soliton_phi_vs_vphi}, we see that there is a one-one correspondence between the separatrix of the phase portrait as shown by a blue curve in figure \ref{P_5_negative_soliton_phi_vs_vphi}(b) with the curve $V(\phi)$ against $\phi$ of figure \ref{P_5_negative_soliton_phi_vs_vphi}(a). 

In figure \ref{P5_Supersoliton_2}, $V(\phi)$ and $\frac{d\phi}{d\xi}$ are plotted against $\phi$ in \ref{P5_Supersoliton_2}(a) and \ref{P5_Supersoliton_2}(b) respectively. In figure \ref{P5_Supersoliton_2}(b), the small solid circles at the points $(0, ~0)$ and $(0.45,~ 0)$ correspond to two saddle points whereas the small solid hexagons at the points $(0.255,~ 0)$ and $(0.69,~ 0)$ correspond to two stable fixed points of the system (\ref{P5_phase_PORTRAITS}). From figure \ref{P5_Supersoliton_2}(b), we see that the separatrix of the phase portrait as shown by the blue curve, which appears to pass through the saddle $(0,~0)$, contains two stable fixed points and another separatrix (green curve), which appears to pass through the non-zero saddle $(0.45,~ 0)$. Therefore, according to Dubinov \& Koltkov \cite{dubinov2012ion} the separatrix (blue curve) which contains at least one separatrix (green curve) that appears to pass through the non-zero saddle $(0.45,~ 0)$ is responsible for the formation of supersoliton. In other words, figure \ref{P5_Supersoliton_2} (a) confirms the existence of supersoliton \cite{dubinov2012ion} if $V(\phi)$ has maximum at some point in the soliton region where $\phi \neq 0$. Again, according to Dubinov \& Koltkov \cite{dubinov2012ion}, if  the differentiation of $\phi$, i.e., the signature of the electric field has at least two maxima (minima), the solitary structure confirms the existence of supersolitons. On the other hand, for conventional soliton, it is simple to check that the differentiation of $\phi$ has only one maximum (minimum). This property has already been explained by Paul \textit{et al.} \cite{paul2018ion}.

From figures \ref{P_5_negative_soliton_phi_vs_vphi} and \ref{P5_Supersoliton_2} we see that each maximum point of $V(\phi)$ generates a saddle point whereas each minimum point of $V(\phi)$ generates a stable equilibrium point of the system (\ref{P5_phase_PORTRAITS}). It is simple to check that in general each maximum value of $V(\phi)$ generates a saddle point whereas each minimum value of $V(\phi)$ generates a stable equilibrium point of the system (\ref{P5_phase_PORTRAITS}). Again, it is important to note that the origin $(0, ~0)$ is always a saddle point of system (\ref{P5_phase_PORTRAITS}) and a separatrix corresponding to a solitary structure appears to pass through the saddle point $(0, ~0)$. The separatrix corresponding to a solitary structure is shown by a blue curve in figures \ref{P_5_negative_soliton_phi_vs_vphi}(b) and \ref{P5_Supersoliton_2}(b) whereas the other separatrix (if exist) is shown by a green curve.
 
Finally, we note that the closed curve about a stable equilibrium point contained in at least one separatrix indicates the possibility of the periodic wave solution about that fixed point. For example, the closed curves (red curves) of figure \ref{P_5_negative_soliton_phi_vs_vphi}(b) about the fixed point $(-0.77,~ 0)$ lying within the separatrix (blue curve) indicate the possibility of the periodic wave solutions about the fixed point $(-0.77, ~0)$. Figure \ref{P_5_negative_soliton_phi_vs_vphi}(a) shows the existence of a negative potential solitary wave, and figure \ref{P_5_negative_soliton_phi_vs_vphi}(b) shows the corresponding phase portrait contains only one saddle point $(0,~ 0)$ and a non-saddle fixed point $(-0.77, ~0)$. Consequently, there exists only one separatrix that appears to pass through the origin enclosing a non-saddle fixed point. More precisely, the trajectory corresponding to the separatrix approaches to the origin as $\xi \to \pm \infty$. It is also important to note that a separatrix corresponding to a solitary structure does not correspond to a periodic solution because for this case, the trajectory takes forever trying to reach a saddle point. In fact, this is the reason that a pseudo-particle associated with the energy integral (\ref{P5_energy_integral}) takes an infinite long time to move away from its unstable position of equilibrium and then it continues its motion until $\phi$ takes the value $\phi_{m}(<0)$, where $V(\phi)=0$ and $V'(\phi)>0$ and again it takes an infinite long time to come back its unstable position of equilibrium \cite{verheest2001waves,paul2017dust}.

%----- FIG-4 --------------
\begin{figure}
\begin{center}
\includegraphics[height=10cm,width=8.5cm]{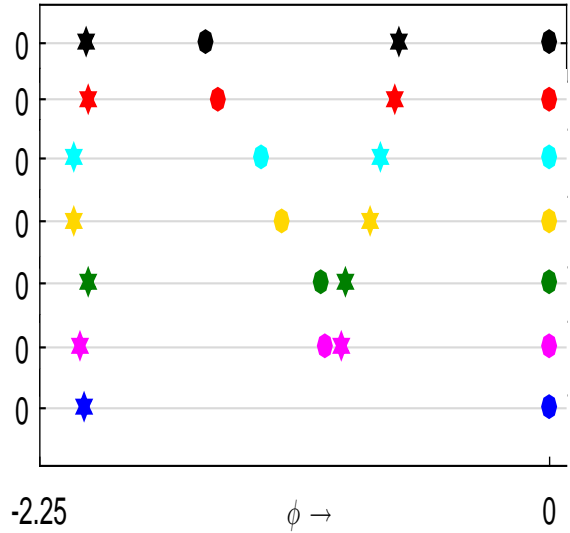}
  \caption{\label{P5_saddle_points} The saddle points (solid circles) and the non-saddle fixed points (solid hexagons) of the system (\ref{P5_phase_PORTRAITS}) have been drawn on the $\phi$-axis for different values of $\cos \theta$ when $n_{sc}=1/9$, $\sigma_{sc}=0.04$, $\beta_{e}=4/103$, $\sigma=0.001$, $\gamma=5/3$, $M=0.98$ and $\omega_{c}=0.2$. Black, red, cyan, yellow, green, magenta and blue colour correspond to $\theta = 39.752^{\circ}$, $\theta = 39.8^{\circ}$, $\theta = 40^{\circ}$, $\theta = 40.1^{\circ}$, $\theta = 40.23^{\circ}$, $\theta = 40.235^{\circ}$ and $\theta = 40.25^{\circ}$.}
\end{center}
\end{figure}

In figure \ref{P5_saddle_points}, we draw the saddle points and stable fixed points of the system (\ref{P5_phase_PORTRAITS}) on the $\phi$-axis for increasing values of $\theta$. From figure \ref{P5_saddle_points}, we have seen that the distance between the non-zero saddle point and a stable fixed point nearest to the origin decreases and ultimately both of them vanish from the system for increasing values of $\theta$. Finally, the system contains only one saddle point at the origin and a stable equilibrium point. Consequently, we have only one separatrix enclosing a stable fixed point and this separatrix appears to pass through the saddle at the origin. Therefore, this separatrix represents a conventional soliton. So, the existence of a soliton after the formation of a double layer confirms the existence of a sequence of supersolitons and there exists a critical value $\theta_{c}$ of $\theta$ such that for the existence of supersolitons after the formation of a double layer, we must have $\theta < \theta_{c}$ whereas for $\theta > \theta_{c}$, we get conventional soliton structures after the formation of a double layer. Thus, Figure \ref{P5_saddle_points} clearly shows the transition of different solitary structures: soliton (before the formation of double layer) $\rightarrow$ double layer $\rightarrow$ supersoliton $\rightarrow$ soliton (soliton after the formation of double layer).

\section{CONCLUSIONS}
In the present paper, we have investigated the existence of arbitrary amplitude ion acoustic solitary waves, double layers and supersoliton structures by considering a collisionless plasma consisting of warm adiabatic ions, Maxwell-Boltzmann distributed cold isothermal electrons and Cairns \cite{cairns1995electrostatic} distributed nonthermal hot electrons immersed in an external uniform static magnetic field ($\textbf{B} = B_{0}\hat{\textbf{z}}$) directed along $z-$axis. But here we have assumed the quasi neutrality condition of charged particulates instead of considering Poisson equation on the basis of the assumption as discussed by Choi \textit{et al.} \cite{choi2005ion}. 

Rufai \textit{et al.} \cite{rufai2016nonlinear} have investigated the finite amplitude IA solitary waves in a collisionless magnetized plasma by considering only one species of Cairns \cite{cairns1995electrostatic} distributed nonthermal electrons whereas in the present paper, we have considered two species of electrons, viz., Maxwell-Boltzmann distributed cold isothermal electrons and Cairns \cite{cairns1995electrostatic} distributed nonthermal hot electrons. In fact, Dalui \textit{et al.} \cite{dalui2017modulational} have extensively discussed the existence of these two different species electrons at different temperatures.

From the expression of $V(\phi)$ as given in equation (\ref{P5_V_phi}), we see that $\displaystyle \frac{V(\phi)}{\omega_{c}^{2}}$ is independent of $\omega_{c}$ and consequently qualitative nature of $V(\phi)$ does not depend on $\omega_{c}$. Therefore, the nature of existence of different solitary structures is independent on $\omega_{c}$.  

We have studied the existence of different solitary structures by considering the variation of $\theta$ and making other parameters of the system as fixed, where $\theta$ is the angle between the direction of the magnetic field and the direction of propagation of the IA wave. But in unmagnetized plasma, Paul \textit{et al.} \cite{paul2017dust}  or in magnetized plasma, Debnath \textit{et al.} \cite{debnath2018ion} have investigated the different solitary structures by considering the variation of Mach number $M$ and making other parameters as fixed.

Again, we have investigated the transition of different solitary structures, viz., soliton (before the formation of double layer) $\rightarrow$ double layer $\rightarrow$ supersoliton $\rightarrow$ soliton (soliton after the formation of double layer), by considering the variation of $\theta$ only and making other parameters of the system as fixed whereas in unmagnetized plasma, Paul \textit{et al.} \cite{paul2017dust}  or in magnetized plasma, Debnath \textit{et al.} \cite{debnath2018ion} have investigated the transition of different solitary structures by considering the variation of Mach number $M$ and making other parameters as fixed. 

%\bibliography{Sandip_5}

\begin{thebibliography}{14}
\expandafter\ifx\csname natexlab\endcsname\relax\def\natexlab#1{#1}\fi
\expandafter\ifx\csname bibnamefont\endcsname\relax
  \def\bibnamefont#1{#1}\fi
\expandafter\ifx\csname bibfnamefont\endcsname\relax
  \def\bibfnamefont#1{#1}\fi
\expandafter\ifx\csname citenamefont\endcsname\relax
  \def\citenamefont#1{#1}\fi
\expandafter\ifx\csname url\endcsname\relax
  \def\url#1{\texttt{#1}}\fi
\expandafter\ifx\csname urlprefix\endcsname\relax\def\urlprefix{URL }\fi
\providecommand{\bibinfo}[2]{#2}
\providecommand{\eprint}[2][]{\url{#2}}

\bibitem[{\citenamefont{Sagdeev and Leontovich}(1966)}]{sagdeev1966reviews}
\bibinfo{author}{\bibfnamefont{R.~Z.} \bibnamefont{Sagdeev}} \bibnamefont{and}
  \bibinfo{author}{\bibfnamefont{M.~A.} \bibnamefont{Leontovich}},
  \emph{\bibinfo{title}{Reviews of Plasma Physics}}, vol.~\bibinfo{volume}{4}
  (\bibinfo{publisher}{Consultants Bureau New York}, \bibinfo{year}{1966}).

\bibitem[{\citenamefont{Washimi and Taniuti}(1966)}]{washimi1966propagation}
\bibinfo{author}{\bibfnamefont{H.}~\bibnamefont{Washimi}} \bibnamefont{and}
  \bibinfo{author}{\bibfnamefont{T.}~\bibnamefont{Taniuti}},
  \bibinfo{journal}{Phys. Rev. Lett.} \textbf{\bibinfo{volume}{17}},
  \bibinfo{pages}{996} (\bibinfo{year}{1966}).

\bibitem[{\citenamefont{Yu et~al.}(1980)\citenamefont{Yu, Shukla, and
  Bujarbarua}}]{yu1980fully}
\bibinfo{author}{\bibfnamefont{M.~Y.} \bibnamefont{Yu}},
  \bibinfo{author}{\bibfnamefont{P.~K.} \bibnamefont{Shukla}},
  \bibnamefont{and}
  \bibinfo{author}{\bibfnamefont{S.}~\bibnamefont{Bujarbarua}},
  \bibinfo{journal}{Phys. Fluids} \textbf{\bibinfo{volume}{23}},
  \bibinfo{pages}{2146} (\bibinfo{year}{1980}).

\bibitem[{\citenamefont{Choi et~al.}(2005)\citenamefont{Choi, Ryu, Lee, and
  Lee}}]{choi2005ion}
\bibinfo{author}{\bibfnamefont{C.~R.} \bibnamefont{Choi}},
  \bibinfo{author}{\bibfnamefont{C.~M.} \bibnamefont{Ryu}},
  \bibinfo{author}{\bibfnamefont{N.~C.} \bibnamefont{Lee}}, \bibnamefont{and}
  \bibinfo{author}{\bibfnamefont{D.~Y.} \bibnamefont{Lee}},
  \bibinfo{journal}{Phys. Plasmas} \textbf{\bibinfo{volume}{12}},
  \bibinfo{pages}{022304} (\bibinfo{year}{2005}).

\bibitem[{\citenamefont{Dalui et~al.}(2017{\natexlab{a}})\citenamefont{Dalui,
  Bandyopadhyay, and Das}}]{dalui2017modulational_2}
\bibinfo{author}{\bibfnamefont{S.}~\bibnamefont{Dalui}},
  \bibinfo{author}{\bibfnamefont{A.}~\bibnamefont{Bandyopadhyay}},
  \bibnamefont{and} \bibinfo{author}{\bibfnamefont{K.~P.} \bibnamefont{Das}},
  \bibinfo{journal}{Phys. Plasmas} \textbf{\bibinfo{volume}{24}},
  \bibinfo{pages}{102310} (\bibinfo{year}{2017}{\natexlab{a}}).

\bibitem[{\citenamefont{Cairns et~al.}(1995)\citenamefont{Cairns, Mamum,
  Bingham, Bostr{\"o}m, Dendy, Nairn, and Shukla}}]{cairns1995electrostatic}
\bibinfo{author}{\bibfnamefont{R.~A.} \bibnamefont{Cairns}},
  \bibinfo{author}{\bibfnamefont{A.~A.} \bibnamefont{Mamum}},
  \bibinfo{author}{\bibfnamefont{R.}~\bibnamefont{Bingham}},
  \bibinfo{author}{\bibfnamefont{R.}~\bibnamefont{Bostr{\"o}m}},
  \bibinfo{author}{\bibfnamefont{R.~O.} \bibnamefont{Dendy}},
  \bibinfo{author}{\bibfnamefont{C.~M.~C.} \bibnamefont{Nairn}},
  \bibnamefont{and} \bibinfo{author}{\bibfnamefont{P.~K.}
  \bibnamefont{Shukla}}, \bibinfo{journal}{Geophys. Res. Lett.}
  \textbf{\bibinfo{volume}{22}}, \bibinfo{pages}{2709} (\bibinfo{year}{1995}).

\bibitem[{\citenamefont{Rufai et~al.}(2014)\citenamefont{Rufai, Bharuthram,
  Singh, and Lakhina}}]{rufai2014ion}
\bibinfo{author}{\bibfnamefont{O.~R.} \bibnamefont{Rufai}},
  \bibinfo{author}{\bibfnamefont{R.}~\bibnamefont{Bharuthram}},
  \bibinfo{author}{\bibfnamefont{S.~V.} \bibnamefont{Singh}}, \bibnamefont{and}
  \bibinfo{author}{\bibfnamefont{G.~S.} \bibnamefont{Lakhina}},
  \bibinfo{journal}{Phys. Plasmas} \textbf{\bibinfo{volume}{21}},
  \bibinfo{pages}{082304} (\bibinfo{year}{2014}).

\bibitem[{\citenamefont{Rufai et~al.}(2016)\citenamefont{Rufai, Bharuthram,
  Singh, and Lakhina}}]{rufai2016nonlinear}
\bibinfo{author}{\bibfnamefont{O.}~\bibnamefont{Rufai}},
  \bibinfo{author}{\bibfnamefont{R.}~\bibnamefont{Bharuthram}},
  \bibinfo{author}{\bibfnamefont{S.~V.} \bibnamefont{Singh}}, \bibnamefont{and}
  \bibinfo{author}{\bibfnamefont{G.~S.} \bibnamefont{Lakhina}},
  \bibinfo{journal}{Phys. Plasmas} \textbf{\bibinfo{volume}{23}},
  \bibinfo{pages}{032309} (\bibinfo{year}{2016}).

\bibitem[{\citenamefont{Debnath et~al.}(2018)\citenamefont{Debnath,
  Bandyopadhyay, and Das}}]{debnath2018ion}
\bibinfo{author}{\bibfnamefont{D.}~\bibnamefont{Debnath}},
  \bibinfo{author}{\bibfnamefont{A.}~\bibnamefont{Bandyopadhyay}},
  \bibnamefont{and} \bibinfo{author}{\bibfnamefont{K.~P.} \bibnamefont{Das}},
  \bibinfo{journal}{Phys. Plasmas} \textbf{\bibinfo{volume}{25}},
  \bibinfo{pages}{033704} (\bibinfo{year}{2018}).

\bibitem[{\citenamefont{Dubinov and Kolotkov}(2012)}]{dubinov2012ion}
\bibinfo{author}{\bibfnamefont{A.~E.} \bibnamefont{Dubinov}} \bibnamefont{and}
  \bibinfo{author}{\bibfnamefont{D.~Y.} \bibnamefont{Kolotkov}},
  \bibinfo{journal}{Plasma Phys. Rep.} \textbf{\bibinfo{volume}{38}},
  \bibinfo{pages}{909} (\bibinfo{year}{2012}).

\bibitem[{\citenamefont{Paul and Bandyopadhyay}(2018)}]{paul2018ion}
\bibinfo{author}{\bibfnamefont{A.}~\bibnamefont{Paul}} \bibnamefont{and}
  \bibinfo{author}{\bibfnamefont{A.}~\bibnamefont{Bandyopadhyay}},
  \bibinfo{journal}{Indian J. Phys.} \textbf{\bibinfo{volume}{92}},
  \bibinfo{pages}{1187} (\bibinfo{year}{2018}).

\bibitem[{\citenamefont{Verheest}(2001)}]{verheest2001waves}
\bibinfo{author}{\bibfnamefont{F.}~\bibnamefont{Verheest}},
  \emph{\bibinfo{title}{Waves in dusty space plasmas}}, vol.
  \bibinfo{volume}{245} (\bibinfo{publisher}{Springer Science \& Business
  Media}, \bibinfo{year}{2001}).

\bibitem[{\citenamefont{Paul et~al.}(2017)\citenamefont{Paul, Bandyopadhyay,
  and Das}}]{paul2017dust}
\bibinfo{author}{\bibfnamefont{A.}~\bibnamefont{Paul}},
  \bibinfo{author}{\bibfnamefont{A.}~\bibnamefont{Bandyopadhyay}},
  \bibnamefont{and} \bibinfo{author}{\bibfnamefont{K.~P.} \bibnamefont{Das}},
  \bibinfo{journal}{Phys. Plasmas} \textbf{\bibinfo{volume}{24}},
  \bibinfo{pages}{013707} (\bibinfo{year}{2017}).

\bibitem[{\citenamefont{Dalui et~al.}(2017{\natexlab{b}})\citenamefont{Dalui,
  Bandyopadhyay, and Das}}]{dalui2017modulational}
\bibinfo{author}{\bibfnamefont{S.}~\bibnamefont{Dalui}},
  \bibinfo{author}{\bibfnamefont{A.}~\bibnamefont{Bandyopadhyay}},
  \bibnamefont{and} \bibinfo{author}{\bibfnamefont{K.~P.} \bibnamefont{Das}},
  \bibinfo{journal}{Phys. Plasmas} \textbf{\bibinfo{volume}{24}},
  \bibinfo{pages}{042305} (\bibinfo{year}{2017}{\natexlab{b}}).

\end{thebibliography}

\end{document}